%% file: main.tex
\documentclass[conference]{IEEEtran}
\IEEEoverridecommandlockouts
\usepackage{cite}
\usepackage{amsmath,amssymb,amsfonts}
\usepackage{graphicx}
\usepackage{textcomp}
\usepackage{xcolor}
\usepackage[hyphens]{url}
\usepackage{braket}
\usepackage{algorithm}
\usepackage{algpseudocode}
\usepackage{multirow}
\usepackage{makecell}
\usepackage{subfig}

\usepackage{blkarray}
\usepackage{widetext}
\usepackage[affil-it]{authblk}

\def\BibTeX{{\rm B\kern-.05em{\sc i\kern-.025em b}\kern-.08em
    T\kern-.1667em\lower.7ex\hbox{E}\kern-.125emX}}
    
\definecolor{backcolour}{rgb}{1,1,1}
\definecolor{myblue}{RGB}{27,95,166}
\definecolor{commentcolour}{rgb}{0.2,0.2,0.6}
\definecolor{codegray}{rgb}{0.5,0.5,0.5}
\definecolor{codepurple}{rgb}{0.58,0,0.82}
\usepackage{listings}

\lstdefinelanguage{Scaffold}{%
  language     = C,
  morekeywords = {module, qbit, Allocate, Compute, Store, Uncompute, Free},
}
\lstdefinestyle{mystyle}{
    backgroundcolor=\color{backcolour},   
    commentstyle=\color{codegray},
    keywordstyle={\bfseries\color{myblue}},
    numberstyle=\tiny\color{codegray},
    stringstyle=\color{codepurple},
    basicstyle=\ttfamily\scriptsize,
    breakatwhitespace=false,         
    breaklines=true,
    mathescape=true,
    captionpos=b,                    
    keepspaces=true,                 
    numbers=left,                    
    numbersep=5pt,                  
    showspaces=false,                
    showstringspaces=false,
    showtabs=false,                  
    tabsize=2,
    frame = single
}
\begin{document}

\title{SQUARE: Strategic Quantum Ancilla Reuse for Modular Quantum Programs\\ via Cost-Effective Uncomputation
}

\author[1]{Yongshan Ding\thanks{Corresponding author: yongshan@uchicago.edu}}
\author[1]{Xin-Chuan Wu}
\author[1,2]{Adam Holmes}
\author[1]{Ash Wiseth}
\author[1]{\\Diana Franklin}
\author[3]{Margaret Martonosi}
\author[1]{Frederic T. Chong}

\affil[1]{Department of Computer Science, University of Chicago, Chicago, IL 60615, USA}
\affil[2]{Intel Labs, Intel Corporation, Hillsboro, OR 97124, USA}
\affil[3]{Department of Computer Science, Princeton University, Princeton, NJ 08544, USA}


\maketitle

\begin{abstract}
Compiling high-level quantum programs to machines that are size constrained (i.e. limited number of quantum bits) and time constrained (i.e. limited number of quantum operations) is challenging. In this paper, we present SQUARE (Strategic QUantum Ancilla REuse), a compilation infrastructure that tackles allocation and reclamation of scratch qubits (called ancilla) in modular quantum programs. At its core, SQUARE strategically performs uncomputation to create opportunities for qubit reuse.

Current Noisy Intermediate-Scale Quantum (NISQ) computers and forward-looking Fault-Tolerant (FT) quantum computers have fundamentally different constraints such as data locality, instruction parallelism, and communication overhead. Our heuristic-based ancilla-reuse algorithm balances these considerations and fits computations into resource-constrained NISQ or FT quantum machines, throttling parallelism when necessary. To precisely capture the workload of a program, we propose an improved metric, the ``active quantum volume,'' and use this metric to evaluate the effectiveness of our algorithm. Our results show that SQUARE improves the average success rate of NISQ applications 
by 1.47X. Surprisingly, the additional gates for uncomputation create ancilla with better locality, and result in substantially fewer swap gates and less gate noise overall. SQUARE also achieves an average reduction of 1.5X (and up to 9.6X)
in active quantum volume for FT machines.
\end{abstract}

\begin{IEEEkeywords}
quantum computing, compiler optimization, reversible logic synthesis
\end{IEEEkeywords}

\input{introduction.tex}
\input{background.tex}
\input{uncompute.tex}

\input{idea_and_motivation.tex}

\input{implementation.tex}
\input{evaluation.tex}

\input{related.tex}

\input{conclusion.tex}

\section*{Acknowledgment}

This work is funded in part by EPiQC, an NSF Expedition in Computing, under grants CCF-1730449/1730082; in part by STAQ, under grant NSF Phy-1818914; and in part by DOE grants DE-SC0020289 and DE-SC0020331. We thank Kenneth Brown, Isaac Chuang, Yipeng Huang, Ali Javadi-Abhari, and Frank Mueller for valuable discussions. We also thank the anonymous reviewers for their useful comments.

\bibliographystyle{IEEEtran}
\bibliography{main.bib}

\end{document}

%% file: introduction.tex
\section{Introduction}\label{sec:intro}
Thanks to recent rapid advances in physical implementation technologies, quantum computing (QC) is seeing an exciting surge of hardware prototypes from both academia and industry \cite{debnath2016demonstration, Google, IBM, Intel}.  This phase of QC development is commonly referred as the Noisy Intermediate-Scale Quantum (NISQ) era \cite{preskill2018quantum}. Current quantum computers are able to perform on the order of hundreds of quantum operations (gates) using tens to hundreds of quantum bits (qubits). While modest in scale, these NISQ machines are large and reliable enough to perform some computational tasks. Looking beyond the NISQ era, quantum computers will ultimately arrive at the Fault-Tolerant (FT) era \cite{bennett1996mixed,gottesman2010introduction}, where quantum error correction is implemented to ensure operation fidelity is met for arbitrarily large computations. 

One major challenge, however, facing the QC community, is the substantial
resource gap between what quantum computer hardware can offer and what quantum algorithms (for classically intractable problems) typically require. Space and time resources in a quantum computer are extremely constrained.
Space is constrained in the sense that there will be a limited number of qubits available, often further complicated by poor connectivity between qubits. Time is also constrained because qubits suffer from decoherence noise and gate noise. Too many successive operations on qubits results in lower program success rates. 

Due to space and time constraints, it is critical to find efficient ways to compile large programs into programs (circuits) that minimize the number of qubits and sequential operations (circuit depth). Several options have been proposed \cite{bertels2019quantum,campbell2017unified,chong2017programming,heyfron2018efficient,paler2017fault,steiger2018projectq, wecker2014liqui}. Among the options, one approach not yet well studied is to coordinate allocation and reclamation of qubits for optimal reuse and load balancing \cite{soeken2018programming}.
Reclaiming qubits, however, comes with a substantial operational cost. In particular, to obey the rules of quantum computation, before recycling a used qubit, additional gate operations need to be applied to ``undo'' part of its computation.  

In this paper, we propose the first automated compilation framework for such strategic quantum ancilla reuse (SQUARE) in modular quantum programs that could be readily applied to both NISQ and FT machines. SQUARE is a compiler that automatically determines places in a program to perform such \emph{uncomputation} in order to manage the trade-offs in qubit savings and gate costs and to optimize for the overall resource consumption.

\begin{figure}[t!]
    \centering
    \includegraphics[width=0.85\linewidth]{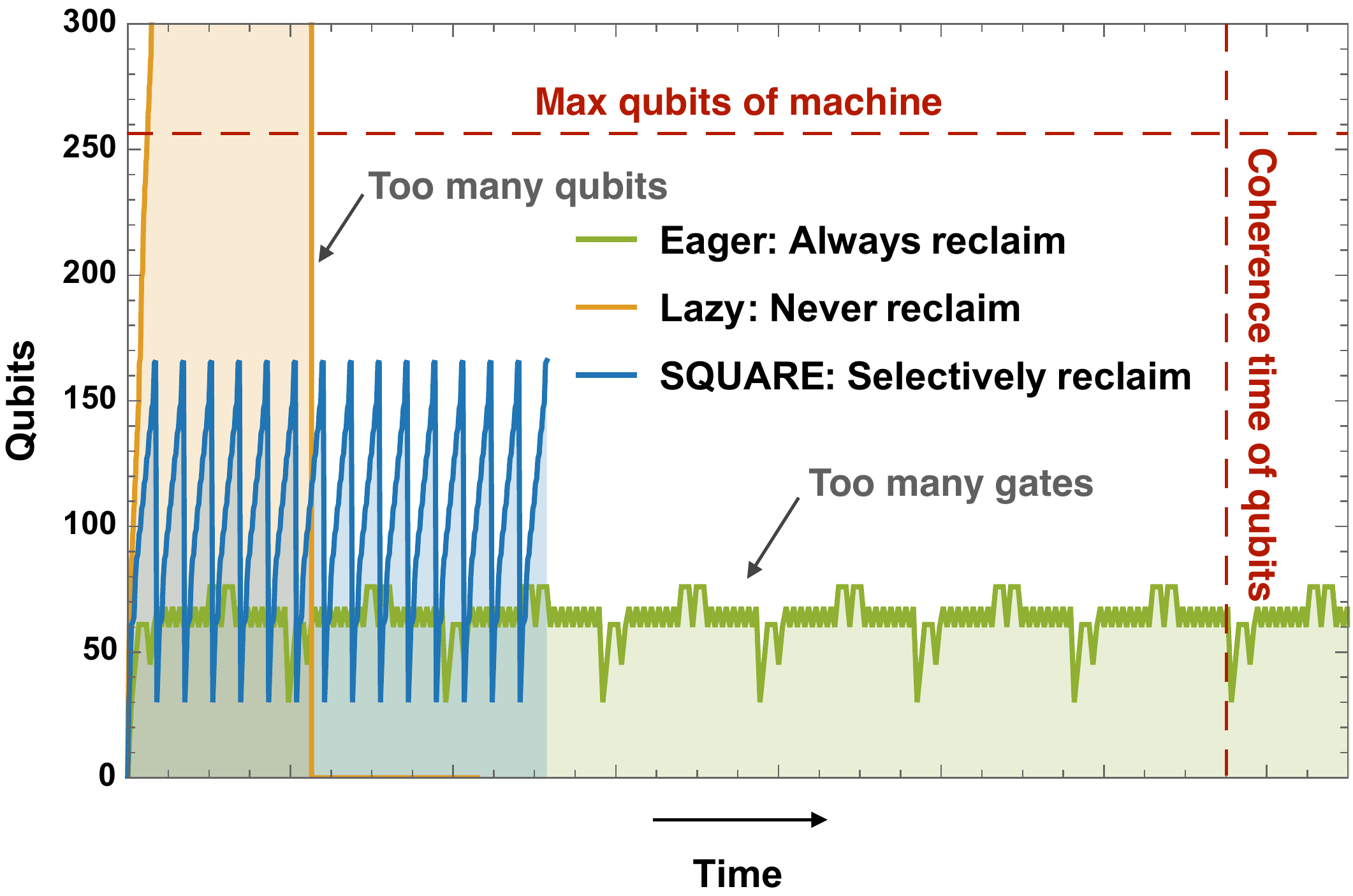}
    \caption{Qubit usage over time for Modular Exponentiation.
The shaded area under the curve corresponds to
the active quantum volume of this application.  The blue curve,
representing a balance between qubit reclamation and uncomputation,
has the lowest area and is the best option.}
    \label{fig:mod_exp_intro}
\end{figure}

Optimally choosing reclamation points in a 
program is crucial in minimizing resource consumption. This is because reclaiming too often can result in significant time
cost (due to more gates dedicated to uncomputation). Likewise, reclaiming too
seldom may require too many qubits (e.g. fail to fit
the program in the machine).  For example, Figure~\ref{fig:mod_exp_intro} shows how 
qubit usage changes over time for the modular exponentiation step in Shor's 
algorithm \cite{shor1999polynomial}. Unfortunately, finding the optimal points in a program for reclaiming qubit could get extremely complex \cite{bennett1989time, knill1995analysis}. An efficient qubit reuse strategy will play a pivotal role in enabling the execution of programs on resource-constrained machines.

To precisely estimate the workload of a computational task,
we propose a resource metric called ``\emph{active quantum volume}'' (AQV) that evaluates the ``liveness'' of qubit
during the lifetime of the program, which we will formally introduce in Section \ref{subsec:aqv}. This is inspired by the concept
of ``quantum volume''  introduced by IBM \cite{bishop2017quantum}, a common measure for the computational capability of a quantum hardware device, based on parameters such as number of qubits,
number of gates, and error probability. AQV is a
metric that measures the volume of resource required by a program when executing on a target hardware, which can therefore serve as an minimization objective for the allocation and reclamation strategies.

The contributions of our work are:
\begin{itemize}
    \item We present a heuristic-based compilation framework, called SQUARE, for optimizing qubit allocation and reclamation in modular reversible programs. It leverages the knowledge of qubit locality as well as program modularity and parallelism.
    \item We introduce a resource metric, active quantum volume (AQV), that calculates the ``liveness'' of qubits over the lifetime of a program. This new metric allows us to quantify the effectiveness of various optimization strategies, as well as to characterize the volume of resources consumed by different computational tasks.
    \item Our approach fits computations into resource-constrained NISQ machines by strategically reusing qubits. Surprisingly, adding gates for uncomputation can \emph{improve} the fidelity of a program rather than impair it, as it creates ancilla with better locality, leading to substantially fewer swap gates and thus less gate noise. SQUARE improves the success rate of NISQ applications by 1.47X on average.
    \item Our approach has broad applicability from NISQ to FT machines. SQUARE achieves an average reduction of 1.5X (and up to 9.6X) in active quantum volume for FT systems.
\end{itemize}

The rest of the paper is organized as follows: Section \ref{sec:bg} briefly discusses the basics of quantum computation and compilation of reversible arithmetic to quantum circuits, as well as related work in both classical and quantum compilation. Section 
\ref{sec:idea} illustrate the central problem of allocation and reclamation of ancilla tackled in this paper and the general idea of our solution. Section \ref{sec:impl} describes in detail the techniques that make up our proposed algorithm. Section \ref{sec:eval} evaluates the performance of the algorithm on an array of benchmarks under the NISQ and FT architectures. Finally, Section \ref{sec:conclude} summarizes and then highlights challenges awaiting satisfactory solutions.


%% file: background.tex
\section{Background and Related Work} \label{sec:bg}

\subsection{Basics of Quantum Computing}
Quantum computers are devices that harness quantum mechanics to store and process information. For this paper, we highlight three of the basic rules derived from the principles of quantum mechanics:

\begin{itemize}
    \item {\it Superposition rule:} A quantum bit (qubit) can be in a quantum state of a linear combination of 0 and 1: $\ket{\psi} = \alpha\ket{0} + \beta\ket{1}$, where $\alpha$ and $\beta$ are complex amplitudes satisfying $|\alpha|^2 + |\beta|^2 = 1$.
    \item {\it Transformation rule:} Computation on qubits is accomplished by applying a unitary quantum logic gate that maps from one quantum state to another. This process is \emph{reversible} and \emph{deterministic}.
    \item {\it Measurement rule:} Measurement or readout of a qubit $\ket{\psi} = \alpha\ket{0} + \beta\ket{1}$ collapses the quantum state to classical outcomes: $\ket{\psi'} = \ket{0}$ with probability $|\alpha|^2$ and $\ket{\psi'} = \ket{1}$ with probability $|\beta|^2$. This is \emph{irreversible} and \emph{probabilistic}.
\end{itemize}

\subsubsection{Reversibility constraints.} 
\begin{figure}[t!]
\begin{center}
\includegraphics[width=0.9\linewidth, trim=-1cm 0 0 0]{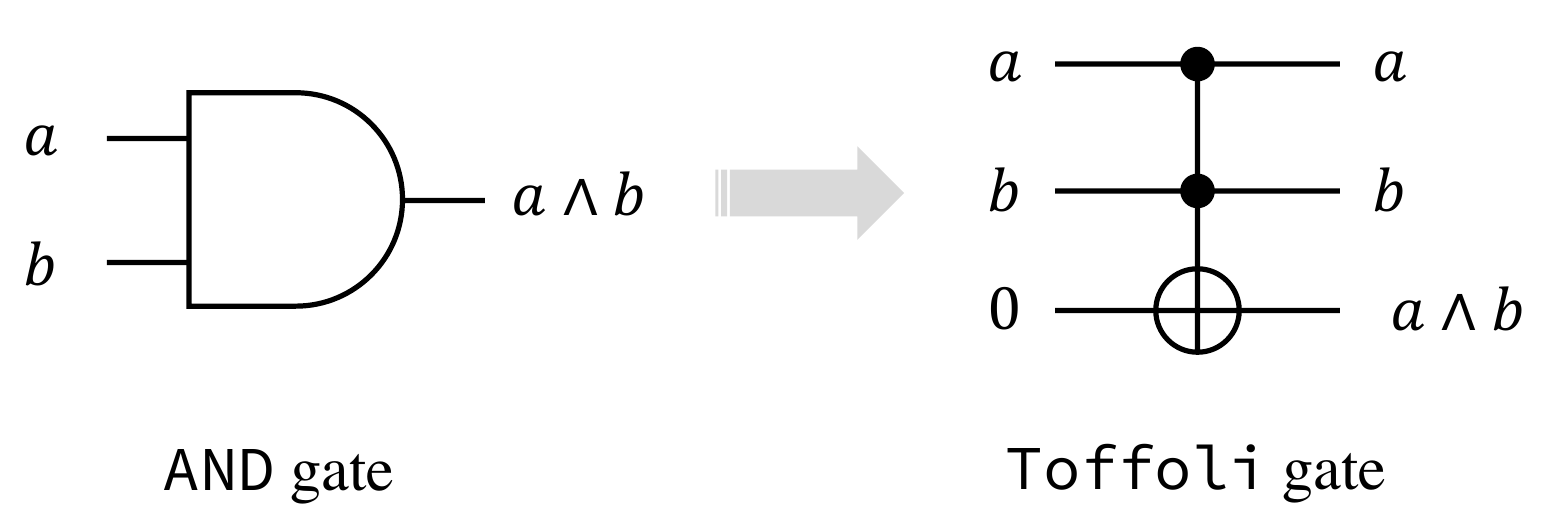}
\caption{Circuit diagram for the irreversible {\tt AND} gate and the reversible {\tt Toffoli} gate.}
\label{fig:and}
\end{center}
\end{figure}

The above three rules give rise to the potential computing power that quantum computers possess, but at the same time, they impose strict constraints on what we can do in quantum computation. For example, the transformation rule implies that any quantum logic gate we apply to a qubit has to be \emph{reversible}. The classical {\tt AND} gate in Figure~\ref{fig:and} is \emph{not} reversible because we cannot recover the two input bits based solely on one output bit. To make it reversible, we could introduce a scratch bit, called \emph{ancilla}, to store the result out-of-place, as in controlled-controlled-{\tt NOT} gate (or {\tt Toffoli} gate) in Figure~\ref{fig:and}. Note that we use the terminology ``ancilla'' in its most general sense--it is not limited to error correction ancilla, but rather, any (physical or logical) qubits used as scratch space for computation.  As the arithmetic complexity scales up when tackling difficult computational problems, we quickly see extensive usage of ancilla bits in our circuits due to this \emph{reversibility constraint}. 

\subsection{Synthesizing Reversible Arithmetic}\label{subsec:rev}


For small arithmetic logic, algorithms exist to directly synthesize reversible circuits from the truth-table of the desired function \cite{grosse2009exact, miller2003transformation, soeken2017hierarchical} and with templates \cite{maslov2005toffoli}. This typically works well for small low-level combinational functions, but not for functions with internal states \cite{parent2015reversible}. As the complexity of the arithmetic in an algorithm scales up, \emph{modularity} quickly becomes convenient, and in many cases necessary. That is, to construct high-level arithmetic, we need to build up from small modular subroutines. 

In reversible logic synthesis and optimization, besides making our circuit for the reversible function contain as few gates as possible, we would also like to minimize the amount of scratch memory (i.e. number of ancilla bits) used in the circuit. Fortunately, there is a way to recycle ancilla bits for later reuse. For a circuit that makes extensive use of scratch memory, managing the allocation and reclamation of the ancilla bits becomes critical to producing an efficient implementation of the function.

\subsubsection{Role of reversible arithmetic in quantum algorithms.} \label{subsec:role}
Reversible arithmetic plays a pivotal role in many known quantum algorithms. The advantage of quantum algorithms is thought to stem from their ability to pass a superposition of inputs into a classical function at once, whereas a classical algorithm can only evaluate the function on single input at a time. Many quantum algorithms involve computing classical functions, which must be embedded in the form of reversible arithmetic subroutines in quantum circuits. For example, Shor's factoring algorithm \cite{shor1999polynomial} uses classical modular-exponentiation arithmetic, Grover's searching algorithm \cite{grover1996fast} also implements its underlying search problem as an oracle subroutine, and the HHL algorithm for solving linear system of equations contains an expensive reciprocal step \cite{harrow2009quantum}. These reversible arithmetic subroutines are typically the most resource-demanding computational components of the entire quantum circuit.

\subsection{Compiling Quantum Circuits to Target Architecture}
As discussed above, there are several options for obtaining a synthesized classical reversible circuit. 
The next step is to compile it down to a sequence of instructions that a quantum machine recognizes and natively supports, that is to resolve \emph{architectural constraints}. This means considering the following two aspects: 

1) {\it Instruction set.} There are certain quantum logic gates that are supported in a given device architecture. In most cases, this gate set is ``Clifford+T'' gates, comprised of the {\tt CNOT} gate, {\tt NOT} gate (or {\tt X} gate), Hadamard gate (or {\tt H} gate) and {\tt T} gate. This is a common set for most of today's gate-based quantum hardware prototypes, as well as for large-scale fault-tolerant machines (e.g. with surface code error correction). Given a classical reversible circuit, we can replace each gate with its quantum counter-part. In particular, {\tt NOT} gates and {\tt CNOT} gates can be directly implemented as quantum gates. For {\tt Toffoli} gates, algorithms exist that decompose them into a sequence of Clifford+T gates \cite{abdessaied2016technology, amy2013meet,kliuchnikov2012fast,maslov2016advantages, welch2014efficient}. At lower level, some instruction sets are proposed to offer direct control over the target hardware \cite{fu2019eqasm}. 
    
2) {\it Qubit communication.} Multi-qubit quantum gates are implemented by interacting the operand qubits with one another. At the physical level, building large-scale quantum machines with all-to-all qubit connectivity is shown to be extremely challenging. The latest effort from IonQ \cite{ionq} offers a machine with 11 fully-connected qubits using trapped-ion technology. Superconducting machines, for instance those by IBM\cite{IBM} and Rigetti\cite{Rigetti}, typically have \emph{much} lower connectivity. Any scalable proposal would involve an architecture of limited qubit connectivity and a model for resolving long-distance interactions. As a consequence, interacting qubits that are not directly connected would induce communication costs. 

\subsubsection{Difference between NISQ and FT machines.} \label{subsubsec:nisqft}
Depending on the topology of the architecture and the model for resolving two-qubit interactions, communication costs will differ. In the context of a NISQ machine, the most frequently used approach to resolve a long-distance two-qubit gate is through swaps, where two (physical) qubits are moved closer by performing a chain of swap gates that connects them. Each \text{SWAP} gate consists of three \texttt{CNOT} gates. The time to complete a swap chain is proportional to the length of the chain. In a FT machine, a logical qubit is encoded by a number of physical qubits. A logical operation is specified by a sequence of physical operations on its physical qubits. For instance, for surface code implementation, physical qubits form a 2D grid with every data qubit connected to its four nearest neighbors through stabilizer ancillas. In essence, a logical operation is defined by specifying how the stabilizer ancillas interact with the data qubits. In particular, a logical two-qubit gate can be defined by braiding\footnote{The focus of this study is on braiding, but other schemes such as lattice surgery \cite{horsman2012surface, litinski2018lattice} exists for resolving two-qubit interactions on surface code, which may expose different communication tradeoffs.} which creates a path between logical qubits, where the stabilzer ancillas along the path do not interact with their neighbors \cite{ding2018magic, javadi2017optimized}. Although it can extend to arbitrary length and shape in constant time, two braids are not allowed to cross. We refer interested readers to \cite{fowler2012surface, gottesman2010introduction} for excellent tutorial.

Although our proposed SQUARE approach is designed to optimize for compiling large quantum algorithms onto medium- to large-scale systems with hundreds or thousands of qubits, we  demonstrate that NISQ machines can benefit from SQUARE optimizations significantly as well. As such, Section~\ref{sec:eval} will include experiments that sweep a large range of system sizes (from tens to thousands of qubits), assuming architectures with their appropriate communication models (e.g. swaps and braiding). The one key difference between swaps and braids is that the time to complete a swap chain is proportional to the length of swap, whereas the time to complete a braid is proportional to the number of crossings with other braids.

 
\subsection{Reclaiming Ancilla Qubits via Uncomputation}\label{sec:reclaim}

Reclaiming qubits is the process of returning them to their original $\ket{0}$ state for future reuse. Due to entanglement, this process could be costly; ancilla qubits that are entangled with data qubits will alter the data qubits' state if they are reset or measured. Fortunately, \emph{uncomputation}, introduced by Bennett \cite{bennett1973logical}, is the process for undoing a computation in order to remove the entanglement relationship between ancilla qubits and data qubits from previous computations. 
Figure \ref{fig:uncompute} (left pane) illustrates this process. In that circuit diagram, the $U_f$ box denotes the circuit that computes a classical function $f$. The garbage produced at the end of $U_f$ is cleaned up by storing the output elsewhere and then undoing the computation. 

This uncomputation approach has two potential limitations: firstly, if uncomputation is not done appropriately, we need to pay for the additional gate cost, and secondly, it only works if the circuit $U_f$ implements classical reversible logic - i.e. can be implemented with {\tt Toffoli} gate alone, optionally with {\tt NOT} gate and {\tt CNOT} gate. Quantum algorithms contain non-classical gates such as Hadamard gate, phase gate and T gate; this work focuses on the part in quantum algorithms that computes classical functions (usually arithmetics) which can be implemented without those gates. As discussed in Section \ref{subsec:role}, classical reversible logic plays a large part of most quantum algorithms.

Related work on optimization of qubit allocation and reclamation in reversible programs dates back to as early as \cite{bennett1989time, buhrman2001time}, where they propose to reduce qubit cost via fine-grained uncomputation at the expense of increasing time. Since then, more \cite{chan2015hardness,frank1999reversibility, knill1995analysis,komarath2018pebbling} have followed in characterizing the complexity of reclamation for programs with complex modular structures. Recent work in \cite{amy2017verified, parent2015reversible} show that knowing the structure of the operations in $U_f$ can also help identify bits that may be eligible for cleanup early. A more recent example \cite{meuli2019reversible} improves the reclamation strategy for straight-line programs using a SAT-based algorithm. Some of the above work emphasizes on identifying reclamation opportunities in a flat program, whereas our focus is on coordinating multiple reclamation points in a larger modular program.

\subsection{Reclaiming Qubits via Measurement and Reset}
If ancilla qubits have already been disentangled from the data qubits, we can directly reclaim them by performing a measurement and reset. We can save the number of qubits, by moving measurements to as early as possible in the program, so early that we can reuse the same qubits after measurement for other computation. Prior art \cite{paler2019faster, paler2016wire} has extensively studied this problem and proposed algorithms for discovering such opportunities.

This measurement-and-reset (M\&R) approach also has limitations: firstly, a near-term challenge for NISQ hardware is to support fast qubit reset. Without it, reusing qubits after measurement could be costly or, in many cases, unfeasible. The state-of-the-art technique for resetting a qubit on a NISQ architecture is by waiting long enough for qubit decoherence to happen naturally, typically on the order of milliseconds for superconducting machines \cite{IBM}, significantly longer than the average gate time around several nanoseconds. FT architectures have much lower (logical) measurement overhead (that is roughly the same as that of a single gate operation), and thus are more amenable to the M\&R approach. Secondly, qubit rewiring as introduced in \cite{paler2016wire} works only if measurements can be done early in a program, which may be rare in quantum algorithms --  measurements are absent in many program (such as arithmetic subroutines) or only present somewhere deep in the circuit. M\&R of a qubit is allowed only after all entangled results are no longer needed, whereas uncomputation can be done partially for any subcircuit. As such, unlike the uncomputation approach, M\&R does not \emph{actively create} qubit reuse opportunities.

%% file: uncompute.tex
\begin{figure*}[t]
    \centering
    \includegraphics[width=0.92\textwidth]{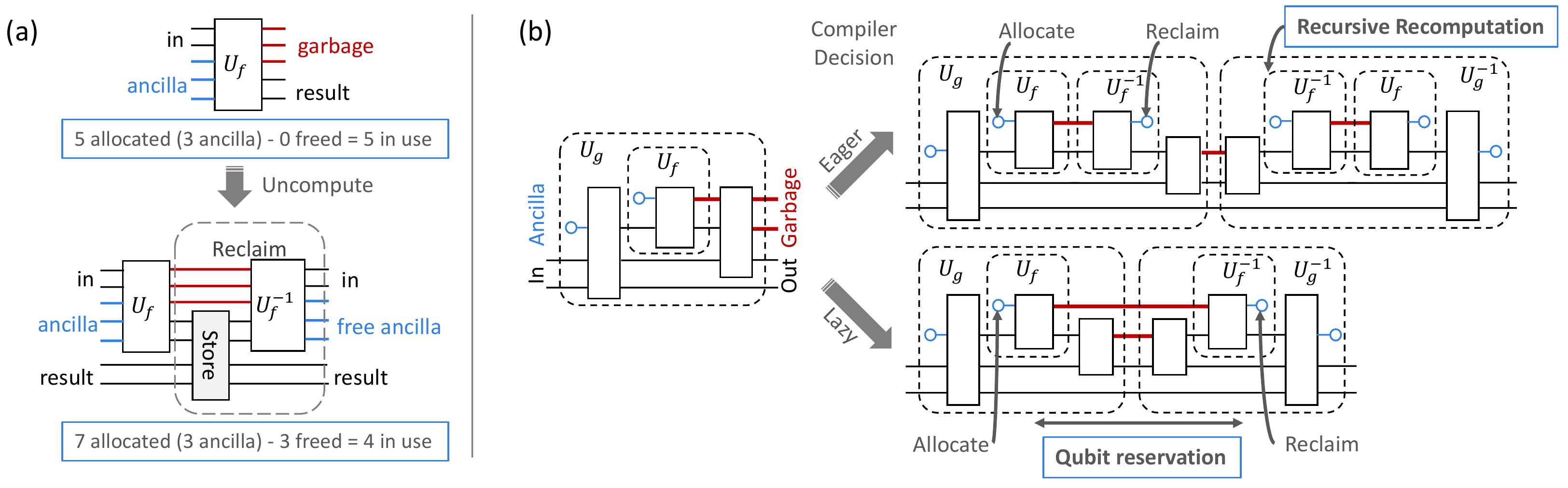}
    \caption{(a) Ancilla qubit reclamation via uncomputation. Each horizontal line is a qubit. Each solid box contains reversible gates. Qubits are highlighted red for the duration of being garbage. (b) Illustration for Eager and Lazy strategies with their respective issues -- recursive recomputation and qubit reservation. Each dashed box denotes a function call containing the enclosed gates. The allocation and reclamation points have been marked as blue circles in the circuit.}
    \label{fig:uncompute} \label{fig:recomp}
\end{figure*}

\subsection{Related Work in Classical Compilation}
Some similarities can be seen in register allocation in classical computing. In that setting, we assign program variables to a limited number of registers in the CPU for fast access. Variables that are not stored in register may be moved to and from RAM, as a process called ``register spilling''. The analysis of live variable and register reuse can be very similar to that of qubits. For instance, our heuristic-based methodology is inspired by register allocation in GPU/distributed systems where communication cost needs to be minimized, and by the technique ``rematerialization'' that reduces the register pressure (i.e. number of registers in use at any point in time) by recomputing some variables instead of storing them to memory. But the trade-offs in qubit allocation and reclamation are unique, which we will introduce as ``recursive recomputation'' and ``qubit reservation'' in Section \ref{subsec:recomp}. Finding the optimal strategy for register allocation, and similarly for qubit reuse, is known to be a hard problem \cite{bouchez2006register, chan2015hardness}.  Luckily, we are able to transfer some general insights from the rich history of classical register allocation optimization to solve the problems in qubit allocation and reclamation. 


The connection made between qubit reuse and classical register allocation \cite{briggs1994improvements, chaitin1982register, poletto1999linear} allows us to inherit some of the intuitions from a wealth of classical literature. Nonetheless, the uncomputation/reuse/locality trade-offs we face are fairly unique. Indeed, rematerialization \cite{briggs1992rematerialization} is very much like qubit reclamation, in that they both aim to lower active registers/qubits at the expense of computation, yet it does not exhibit the same exponential recomputation cost, nor is the increase in the live-range of variables from the recomputing step the same as qubit reservation caused by not uncomputing. We also gained general insights from numerous techniques in code scheduling\cite{goodman2014code, Pinter1993register}, and thread-level parallelism\cite{Xie2015enabling}.

%% file: idea_and_motivation.tex
\section{Key Idea and Motivation}\label{sec:idea}


%

This paper focuses primarily on reusing qubits via uncomputation, and discusses the significance of our proposed strategy in current noisy intermediate-scale quantum (NISQ) and future fault-tolerant (FT) architectures. Prior work such as \cite{parent2015reversible} follows two basic strategies: ``Eager'' cleanup  and ``Lazy'' cleanup, as illustrated in Figure \ref{fig:recomp}.

\vspace{0.2cm}\noindent\textbf{Baseline 1 ``Eager'': Recursive Recomputation.}\label{subsubsec:eager} Eager reclaims qubits at the end of every function. In the example of Figure \ref{fig:recomp}, Eager performs uncomputation at the end of both $U_f$ and $U_g$. When reclaiming ancilla qubits in such programs with nested functions, the uncompute step of the caller would have to repeat \emph{everything} inside of its callee, including the callee's uncompute step. This hierarchical structure will consequently lead to re-computation of the callees, as marked in Figure \ref{fig:recomp}. 
More formally, for a hierarchical program with $\ell$ levels, in the worst case, recomputation causes the number of steps to increase by a factor of $2^\ell$. We call this exponential blowup phenomenon ``recursive recomputation''. That is why the 2-level program in Figure \ref{fig:recomp} has roughly 4 times more steps as the original circuit. This factor will play a crucial role in our heuristic design.

\vspace{0.2cm}\noindent\textbf{Baseline 2 ``Lazy'': Qubit Reservation.}\label{subsubsec:lazy} Lazy reclaims qubits only at the top-level function. In Figure \ref{fig:recomp}, this means only $U_g$ is uncomputed, but not $U_f$. Lazy can sometimes be a preferred strategy because it avoids the wasted recomputation\footnote{There are exceptions, such as recursive Fourier sampling, where recomputation cannot be avoided and is required for correctness \cite{aaronson2003quantum}.}. In other words, it is sometimes beneficial to temporarily leave the garbage of callees, and uncompute the garbage by their callers. This is equivalent to \emph{inlining} the callee into the caller, and letting the caller handle the reclamation of all ancilla qubits. However, with the benefit of the avoided recomputation comes the cost of ``qubit reservation''. The ancilla qubits from callee are \emph{reserved} or \emph{blocked out} from any reuse until the end of the caller. This can be seen at the bottom right of the example in Figure~\ref{fig:recomp}. The garbage qubit from $U_f$ stays as garbage until almost the end of $U_g^{-1}$, whereas in the Eager case, it is cleaned up right away. 

\subsection{Overview of SQUARE Algorithm}\label{sec:overview}

\begin{figure*}[t]
    \centering
    \includegraphics[width=0.65\textwidth]{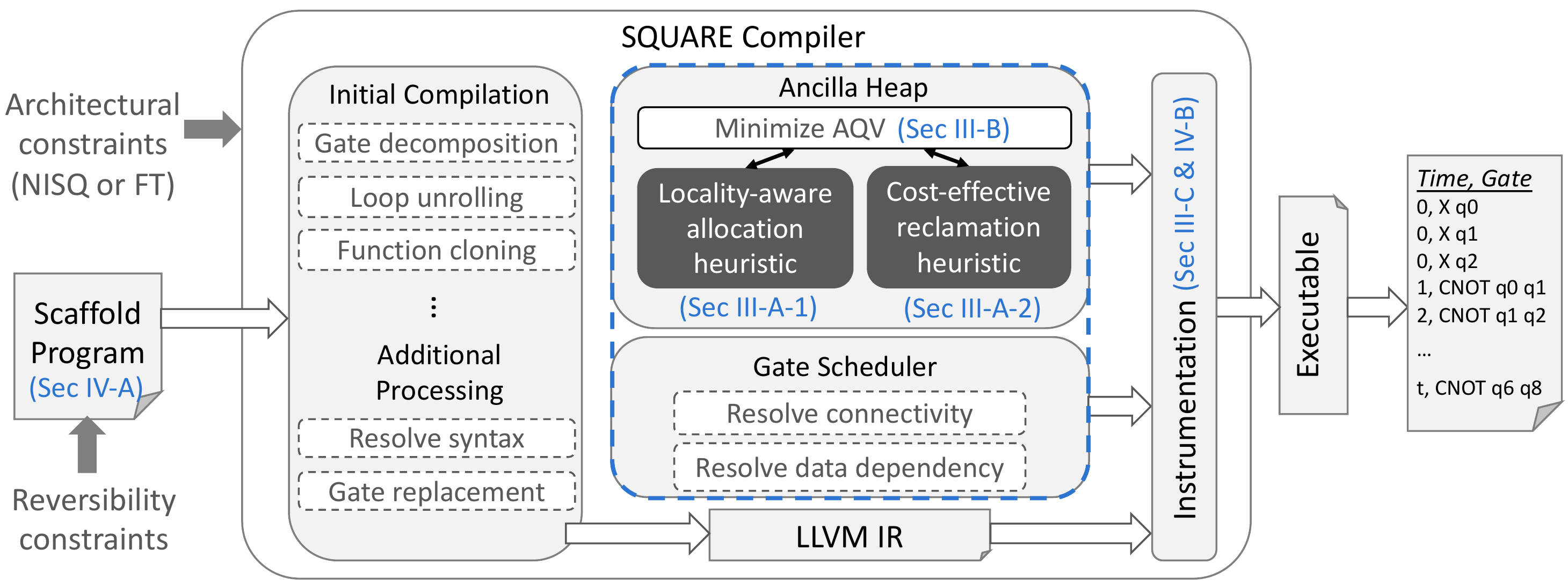}
    \caption{Our Strategic Quantum Ancilla Reuse (SQUARE) compilation flow. SQUARE takes as input a Scaffold \cite{scaffcc} program (see sample code in Figure~\ref{fig:pseudocode}) and produces an executable that simulates the dynamics of qubit allocation/reclamation and gate scheduling, which can then prints out an optimized schedule of quantum gate instructions.}
    \label{fig:flow}
\end{figure*}

\label{subsec:reuse}
Most existing qubit reuse algorithms \cite{bennett1989time, buhrman2001time, parent2015reversible} emphasize on the asymptotic qubit savings, and commonly make an ideal assumption that machines have all-to-all qubit connectivity (i.e. no locality constraint). Since all qubits are considered identical, a straightforward way to keep track of qubits is to maintain a global pool, sometimes referred to as the \emph{ancilla heap}. Ancilla qubits are pushed to the heap when they are reclaimed, and popped off when they are allocated, for instance in a last-in-first-out (LIFO) manner. In this ideal model, we can simply track qubit usage by counting the total number of fresh qubits ever allocated during the lifetime of a program. However, leading proposals of NISQ and FT quantum architectures have far stricter locality constraints. 

\begin{figure}[t]
    \centering
    \includegraphics[width=0.6\linewidth]{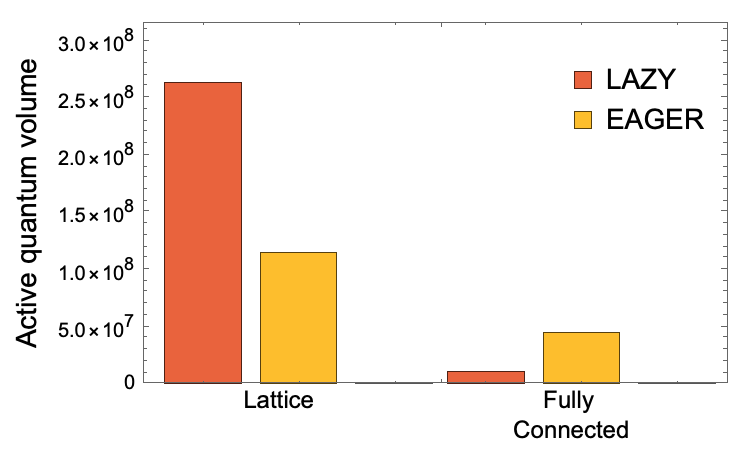}
    \caption{Locality constraint changes the desired reclamation strategy. Results are based on a synthetic benchmark ``Belle''. Lower active quantum volume (defined in Section \ref{subsec:aqv}) is better. Belle performs better on a lattice machine with Eager strategy, while preferring Lazy when operating on a fully-connected machine.}
    \label{fig:swap}
\end{figure}

Our Strategic QUantum Ancilla REuse (SQUARE) algorithm is highly motivated by the lesson that communication can be a determinant factor for qubit allocation and reclamation. Take the NISQ architecture as an example. We make the following two novel observations. Firstly, \emph{same algorithm needs different strategies for different machine connectivity.} In Figure~\ref{fig:swap}, a benchmark named Belle (whose details can be found in Section \ref{subsec:benchmarks}) prefers Eager strategy on a 2-D lattice topology (with swaps), but Lazy strategy on a fully-connected topology (without swaps). Secondly, and most counter-intuitively, \emph{adding uncomputation gates can improve overall circuit fidelity, if done properly}. With careful allocation and reclamation, the expense of additional uncomputation gates is compensated by the reduction of communication cost. This is because uncomputation allows us to create ancilla with better locality, resulting in fewer swap gates and less overall gate noise.

Now we discuss how SQUARE finds the strategies for allocation and reclamation.

\subsubsection{Locality-Aware Allocation (LAA).}\label{subsubsec:laa}


We present the Locality-Aware Allocation (LAA) heuristic in the SQUARE algorithm that prioritizes qubits according to their locations in the machine. At a high level, LAA chooses qubits from the ancilla heap by balancing three main considerations -- communication, serialization, and area expansion. 

When deciding which qubits to allocate and reuse, our heuristic-based algorithm assigns priorities to all qubits. The priorities are weighted not only by the communication overhead of two-qubit interactions but also by their potential impact to the parallelism of the program.  Reusing qubits adds data dependencies to a program and thus serializes computation (which is similar to how reusing register names could lead to false data dependencies and serialization), but not reusing qubits expands the area of active qubits and thus increases the communication overhead between them. Recall from Section \ref{subsubsec:nisqft}, communication have different tradeoffs under NISQ and FT architectures. We will make this distinction in terms of our heuristics clearer in Section \ref{subsec:alloc_detail}.

\subsubsection{Cost-Effective Reclamation (CER).} \label{subsubsec:cer}
The Cost-Effective Reclamation (CER) heuristic makes uncomputation decisions with a simple \emph{cost-benefit analysis}: at each potential reclamation point, we estimate and compare two quantities:
 \begin{itemize}
     \item $C_1$: cost of uncomputation and reclaiming ancilla qubits;
     \item $C_0$: cost of no uncomputation and leaving garbage qubits.
 \end{itemize} 

CER balances the cost of recursive recomputation and qubit reservation as discussed in Section \ref{sec:reclaim}. To do so, we need an efficient way to accurately estimate the $C_1$ and $C_0$ quantities.

In particular, the decision of child function affects not only the cost of itself, but also the cost of parent function. If a child function decides to uncompute, the additional gate costs need to be duplicated should its parent decide to uncompute as well. This was illustrated in Section \ref{subsec:recomp} as the phenomenon we called ``recomputation''. Thus, we should take the level of the function into account when we make the decision. The total cost of a uncomputation, $C_1$, can be expressed as:
 
 \begin{align}\label{eq:c1}
 C_1 = N_{active} \times G_{uncomp} \times S \times 2^\ell
 \end{align}
 where $N_{active}$ is the number of active qubits, $G_{uncomp}$ is the number of gates for uncomputation (including those in all children functions), $\ell$ is the level of the child function in the program call graph, and $S$ is the communication factor. Details can be found in Section \ref{subsec:reclaim_details}.

Now, suppose a function does \emph{not} uncomputing/reclaiming ancilla, the next chance to reclaim them is when its parent function uncomputes. Thus, we want to estimate the cost of holding the ancilla live until the parent's uncompute block is executed. The cost, $C_0$, can be approximated as:
 
 \begin{align}\label{eq:c0}
     C_0 = N_{anc} \times G_p \times S \times \sqrt{(N_{active}+N_{anc})/N_{active}}
 \end{align}
 where $N_{anc}$ is the number of ancillae held by the function, $G_p$ is the number of gates from the current function to the parent's uncompute function. The term under the square root sign captures the effect of `area expansion'', which we will discuss in greater detail in Section \ref{subsec:reclaim_details}.

\subsection{Active Quantum Volume}\label{subsec:aqv}


To accurately estimate the workload of a program, we define the active quantum volume (AQV) of a program as:

\begin{align*}
    V_A = \sum_{q \in Q}\;\; \sum_{(t_{i}, t_{f}) \in T_q} (t_{f} - t_{i})
\end{align*}
where $Q$ is the set of all qubits in the system, and $T_q$ is a sequence of pairs $\{(t_i^0, t_f^0),(t_i^1, t_f^1), \dotsc, (t_i^{|T_q|-1}, t_f^{|T_q|-1})\}$. Each pair corresponds to a qubit usage segment, that is we denote $t_i^k$ and $t_f^k$ as the allocation time and reclamation time of the $k^{th}$ time that qubit $q$ is being used, respectively. AQV is high when a large number of qubits stay ``live'' (in-use) during the execution; thus, the higher the AQV, the more costly it is to execute on that target machine.

The key to this metric is in the term ``active''. In particular, we exclude the time that a reclaimed qubit spends in the heap from volume calculation, because it has been restored to the $\ket{0}$ state (ground state), which does not suffer from the decoherence noise as an excited state does.  Hence, AQV serves as a \emph{minimization objective} in SQUARE. There are a few practical advantages for using AQV over other resource metrics:
\begin{enumerate}
    \item AQV is a better measure of the exposure to errors than the space$\times$time metric (i.e. number of qubits times circuit depth) \cite{fowler2012surface, holmes2019resource}. The more time a qubit stays live, the more susceptible it is to noise from its surroundings. We show lower AQV yields higher success rate in Section \ref{subsec:nisq_exp}.
    \item Unlike qubit count, gate count, or circuit depth, AQV allows us to more accurately model ``liveness'' of qubits on a machine (i.e. which qubits are actively carrying information and performing computation as opposed to staying in ground state unused). \cite{noisy-asplos19} and \cite{tannu2018case} shows that keeping a preferred subset of qubits live can boost program success rate.
    \item IBM's quantum volume (QV) \cite{bishop2017quantum} characterizes the amount of computational resource a quantum device offers, AQV measures the portion of resource being actively utilized by a program on the device. 
\end{enumerate}

\subsection{Compilation Tool Flow of SQUARE}\label{subsec:compile}

In a nutshell, our SQUARE compilation algorithm takes as input a Scaffold program \cite{scaffcc} and produces an optimized schedule of all of its quantum instructions. This is accomplished through what is known as the ``instrumentation-driven'' approach, also used in \cite{heckey-asplos15}, which allows us to pre-simulate the control flow in a quantum program. This works because all inputs are known at compile time for most quantum programs, so we can use their known control flow to simulate resource usage. 

Figure~\ref{fig:flow} illustrates in detail the compilation flow for SQUARE. It consists of three main components: 1) an easy-to-use syntactical construct compatible within the Scaffold language, 2) a qubit allocation heuristic, and 3) a qubit reclamation heuristic.

Under the hood, an input program first goes through an initial compilation step, where each {\tt Allocate()} and {\tt Free()} instruction is replaced by a classical function call (such as in C/C++) that implements the heuristic algorithm. Each quantum gate is replaced by a classical function that resolves the connectivity constraints of its operand qubits and then schedules the gate to the earliest time step possible. As a result, we have obtained an executable for the classical control flow of the quantum program. The compiler maintains an ancilla heap (i.e. pool of reclaimed qubits) that stores all the reclaimed ancilla qubits. Future allocations can therefore choose to pop from the ancilla heap or initialize brand new qubits. One of the key contributions of our work is a heuristic that makes such decisions.

%
%

\label{subsec:recomp}
\subsection{Complexity of SQUARE} 

SQUARE is a heuristic-based greedy algorithm. It makes allocation and reclamation decisions as they appear in program order. As a result, it takes time that scales linearly to the number of reclamation points in a program. Consider a program with nested functions -- all decisions in the callees are made prior to that of the caller, so when deciding for the caller function, the cost of uncomputation is deterministic and easy to estimate. On the flip side, we could end up in a sub-optimal situation where callee's decisions are made neglecting the potential burden for uncomputing its caller. 

The computational complexity of qubit reclamation via uncomputation has been studied. It has been shown that, for programs with linear sequential dependency graph, we can use the reversible pebbling game to approach this problem \cite{meuli2019reversible}. However, finding the optimal points in a program with hierarchical structure is PSPACE complete \cite{chan2015hardness}. For a program with $\ell$ levels and $d$ callees per function, there can be as many as $d^\ell$ possible reclamation points in the worst case. We could be dealing with $2^{d^\ell}$ different combinations of reclamation decisions. So clearly, the na\"ive way for finding the optimal strategy by exhaustively enumerating all possible decisions is far from efficient.

 

%% file: implementation.tex
\section{Implementation Details of SQUARE}\label{sec:impl}

\begin{figure}[t!]
\centering
\begin{minipage}{0.8\linewidth}
\begin{lstlisting}[language=Scaffold]
#include "qalloc.h"

void fun1(qbit* in, qbit* out) {
  qbit anc[1];
  Allocate(anc, 1);
 	Compute {
 	  Toffoli(in[0], in[1], in[2]);
 	  CNOT(in[2], anc[0]);
 	  Toffoli(in[1], in[0], anc[0]);
 	}
 	Store {
 	  CNOT(anc[0], out[0]);
 	}
 	Uncompute{
 	  // Invoke Inverse() to populate
 	  // Or write out explicitly:
 	  Toffoli(in[1], in[0], anc[0]);
 	  CNOT(in[2], anc[0]);
 	  Toffoli(in[0], in[1], in[2]);
 	}
 	Free(anc, 1);
}

int main () {
 	qbit new[4]; // declare name
	Allocate(new, 4); // allocate qubits
 	fun1(new, &new[3]);
 	return 0;
}
\end{lstlisting}
\end{minipage}
\caption{Format of \emph{compute-store-uncompute} construct for qubit allocation and reclamation. Shown here an example function ({\tt fun1}) that allocates and reclaims an ancilla qubit.}
\label{fig:pseudocode}
\end{figure}

In this section, we describe the implementation details of the components of SQUARE algorithm, including the expressive syntactical construct in the Scaffold programming language \cite{scaffcc} that exposes the optimization opportunities, the instrumentation-driven LLVM \cite{heckey-asplos15} that translates the quantum program into a classical executable, and details of the locality-aware allocation heuristics and the cost-effective reclamation heuristics that we left out from Section \ref{subsec:reuse}.

\subsection{Syntactical Construct}\label{subsec:syntax}

In order to express the opportunities for qubit allocation and reclamation optimizations, we augment the high-level Scaffold \cite{scaffcc} programming language with an additional syntactical construct: {\it Compute-Store-Uncompute Code Blocks}. As shown in Figure~\ref{fig:pseudocode}, the keywords ``Allocate'' and ``Free'' are used to express the locations of qubit allocation and reclamation respectively. To enable automation in the optimizations, the compiler needs additional information about the code structure. By  writing a {\tt Compute} code block, the program now has explicitly specified the set of instructions that belong to forward computation. Optionally, programmer can choose to automatically generate the content of the {\tt Uncompute} block by invoking {\tt Inverse()}.

Under the hood, the compiler will replace each {\tt Allocate} and {\tt Free} instruction with an invocation to our heuristic algorithms. Depending on the reclamation decision, it will either execute or skip the uncomputation step accordingly.

\subsection{Instrumentation-Driven Compilation}\label{subsec:instr}

In this section, we illustrate a number of advantages of the instrumentation-drive approach over the conventional pass-drive approach used in most quantum compilers.

The traditional pass-driven approach for compiling and optimizing quantum programs is done by sending a high-level quantum program through multiple layers of transformations, each of which completes a different task. For instance, we have transformations to resolve classical control structures (e.g. loop unrolling and module inlining), explore circuit optimizations (e.g. commutativity and parallelism), satisfy architectural constraints (e.g. qubit connectivity), assign qubit mappings, and perform gate scheduling, etc. One of the potential limitations in this approach is that each transformation performs independently, and in some cases even conflicts with each other \cite{goodman2014code}. So it is very hard to jointly optimize for some correlated problems such as mapping and scheduling. Techniques such as feedback loops could in some cases work well in practice. 

Two main reasons that the instrumentation-driven approach may be a more natural fit for our purpose are: the dynamic nature of our optimization and compilation time scalability. Recall from Section \ref{subsec:compile}, our compilation tool flow produces an executable that allows us to dynamically optimize for the allocation and reclamation of qubits in reversible programs with parallel and modular structures.  In the next section, we illustrate the details of our heuristic algorithms and how they are integrated in the compilation tool flow.

\subsection{Allocation Policy Details} \label{subsec:alloc_detail}
The allocation policy is most concerned about the communication overhead of two-qubit operations in a program. 
\begin{itemize}
    \item Under NISQ architecture, communication between two qubits is accomplished by move one qubit to another via a series of swaps. So swap distance is a direct measure of the locality. The higher the distance, the longer it takes for a chain of swaps to complete.
    \item The concept of locality can be trickier in a FT architecture. Communication is accomplished via braiding. Braids can have arbitrary length or shape, but they are not allowed to cross. As \cite{ding2018magic} shows, average braid length and average braid spacing are both strongly correlated with the number of braid crossings. So we can reduce communication overhead by moving interacting qubits closer and moving non-interacting qubits far apart.
\end{itemize}



When there are fewer qubits available than requested (due to either the maximum qubit constraint or a shortage in the ancilla heap), we mark the allocation as pending, and proceed to schedule all non-dependent, parallel computation and reclamation. Allocation requests are not fulfilled until sufficient ancilla qubits have been reclaimed.

 \begin{table}[t!]
\footnotesize
\centering
\begin{tabular}{cl}
\hline\hline
Algorithm & Description\\\hline
Eager &  \makecell[l]{Reclaim qubits whenever possible, as shown in Section \ref{subsubsec:eager}.} \\\hline
Lazy & \makecell[l]{Only reclaim qubits from the top level in the program \\call graph, as shown in Section \ref{subsubsec:lazy}.} \\\hline
SQUARE & \makecell[l]{Combines Locality-aware allocation (LAA) and \\Cost-effective reclamation (CER). See Section \ref{subsec:reuse}.}\\
\hline\hline
\end{tabular}
\caption{List of compiler configurations.}
\label{tab:config}
\end{table}

 \subsection{Reclamation Policy Details}
\label{subsec:reclaim_details}
The reclamation policy dictates what and when ancilla qubits get recycled. The decisions rely heavily on three main considerations: qubit savings, uncomputation gate count, and communication overhead. In Section \ref{subsubsec:cer}, we have discussed how SQUARE balances between qubit savings and gate count. Now we present further details on how to estimate the communication factor in Equation \ref{eq:c1} and \ref{eq:c0}.
  \begin{itemize}
    \item NISQ architecture: We use the average swap-chain length per gate as the estimate for $S$. This is obtained from the history of swap chains during the compile time simulation -- we keep a running average of the number of swaps for the gates we  scheduled, and use it as an estimate for the subsequent gates in the same module. 
    \item Fault-tolerant (FT) architecture: We use average braiding conflicts per gate as the estimate for $S$. The communication latency due to braid routing is estimated (similarly as \cite{ding2018magic}) by factoring in the average braid length, average braid spacing, and number of crossings. 
\end{itemize}
 
Since ``qubit reservation'' causes the active qubit area to be expanded, leading to higher communication overhead, the multiplicative factor $\sqrt{(N_{active}+N_{anc})/N_{active}}$ aims to estimate the swap or braid length increase due to the expansion.
 

Algorithm \ref{algo:alloc} and \ref{algo:free} are pseudo-code of SQUARE, implementing LAA and CER respectively. Procedures under namespace LLVM are functions that operate on the LLVM IR. In particular, \emph{get\_interact\_qubits()} obtains the set of qubits with which the allocated qubits interact by looking ahead in the code block. \emph{gen\_uncompute\_block()} and \emph{rm\_uncompute\_block()} conditionally expands or deletes the code block under \texttt{Uncompute\{\}} (as shown in Figure \ref{fig:pseudocode}). \emph{closest\_qubit\_in\_heap()} and \emph{closest\_qubit\_new()} look for available qubits to reuse from the heap and from new qubits, respectively. Both functions return the candidate qubits and scores. The scores are calculated based on the communication, serialization, and area expansion, as described in Sec~\ref{subsec:alloc_detail}. We select the qubits with minimum scores until the requested $n$ qubits are allocated.

\begin{algorithm}[H]
\small
 \caption{Allocate: \emph{Locality-Aware Allocation}}
 \label{algo:alloc}
 \textbf{Input:} Number of qubits $n$\\
\textbf{Output:} Set of qubits $\mathcal{S}$
\begin{algorithmic}[1]
\State $\mathcal{I} \gets$ LLVM::get\_interact\_qubits()
\State $\mathcal{S} \gets \emptyset$;
 \For{$i \gets 1$ \textbf{to} $n$}
  \State $q_1, score_1 \gets$ closest\_qubit\_in\_heap($\mathcal{I}$)
  \State $q_2, score_2 \gets$ closest\_qubit\_new($\mathcal{I}$)
  \If{$score_1 \leq score_2$}
   \State $\mathcal{S} \gets \mathcal{S} \cup \{q_1\}$\;
   \Else
   \State $\mathcal{S} \gets \mathcal{S} \cup \{q_2\}$\;
   \EndIf
 \EndFor
 \end{algorithmic}
\end{algorithm}

\begin{algorithm}[H]
\small
 \caption{Free: \emph{Cost-Effective Reclamation}}
  \label{algo:free}
\textbf{Input:} Number of qubits $n$, Set of qubits $\mathcal{S}$
\begin{algorithmic}[1]
\State $C_1 \gets \text{cost of uncomputation}$
\State $C_0 \gets \text{cost of no uncomputation}$
  \If{$C_1 \leq C_0$}
   \State LLVM::gen\_uncompute\_block()\;
   \State heap\_push($n, \mathcal{S}$)\;
   \Else
   \State LLVM::rm\_uncompute\_block()\;
   \State LLVM::transfer\_to\_parent($n, \mathcal{S}$)\;
   \EndIf
 \end{algorithmic}
\end{algorithm}

%% file: evaluation.tex
\section{Evaluation}\label{sec:eval}

\subsection{Benchmarks}\label{subsec:benchmarks}
Table \ref{tab:benchmarks} lists the QC benchmarks and brief description in our study. These benchmarks are reversible arithmetic functions or applications that use ancilla qubits. Since ancilla qubits are expensive in both NISQ and FT architectures, it is crucial to reuse ancilla qubits and improve the success rate of a program.  The first 4 benchmarks (RD53, 6SYM, 2OF5, and ADDER4) are small arithmetic functions suitable for executing on NISQ systems ($10$ - $100$ qubits). The rest of the benchmarks are medium to large functions that are more demanding in computational resources than current NISQ systems can offer. The number of qubits they use, for instance, is on the order of hundreds or thousands. For the last three benchmarks, we construct random synthetic circuits (Jasmine, Elsa, and Belle) with different characteristics in their program structures. In particular, a benchmark is parameterized by the size and shape of its program call graph using 5 variables:  number of nested levels, max number of callees per function, max number of input qubits per function, max number of ancilla qubits per function, maximum number of gates per function.

\begin{table}[t!]
\footnotesize
\centering
\begin{tabular}{cl}
\hline\hline
Name & Description\\\hline
RD53 & Input weight function with 5 inputs and 3 outputs. \\
6SYM & Function with 6 inputs and 1 output. \\
2OF5 & Output is 1 if number of 1s in its input equals two. \\
ADDER4 &  4-bit in-place controlled-addition\footnotemark. \\
Jasmine/Elsa/Belle-s & Small and shallow instances of synthetic benchmarks.\\\hline
ADDER32 &  32-bit in-place controlled-addition. \\
ADDER64 &  64-bit in-place controlled-addition. \\
MUL32 &  32-bit out-of-place controlled-multiplier. \\
MUL64 &  64-bit out-of-place controlled-multiplier. \\
MODEXP &  Modular exponentiation function\footnotemark. \\
SHA2 &  Cryptographic hash function\footnotemark. \\
SALSA20 &  Stream cipher core function\footnotemark. \\
Jasmine &  Shallowly nested synthetic function\footnotemark. \\
Elsa & Heavy workload and shallowly nested synthetic function. \\
Belle & Light workload and deeply nested synthetic function. \\
\hline\hline
\end{tabular}
\caption{Characteristics of benchmark programs.}
\label{tab:benchmarks}
\end{table}

\addtocounter{footnote}{-5}
\stepcounter{footnote}\footnotetext{The adders are based on the Cucarro adder \cite{cuccaro2004new, markov2012constant}.}
\stepcounter{footnote}\footnotetext{Modular exponentiation is an important subroutine used in Shor's factoring algorithm\cite{shor1999polynomial}.}
\stepcounter{footnote}\footnotetext{SHA2 contains multiple rounds of in-place modular additions and bit rotations, based on the implementation from \cite{parent2015reversible}. When used as an oracle in Grover's algorithm\cite{grover1996fast}, we can find hash collisions more efficiently, and thereby reduce the security of the hash function.}

\stepcounter{footnote}\footnotetext{Salsa20 involves 20 rounds of 4 parallel modules. Each module modifies 4 words with modular additions, XOR operations, and bit rotations. 
The Salsa20 stream cipher uses the Salsa20 core function to encrypt data. \cite{bernstein2008salsa20} Salsa family functions have been popularly adopted for TLS in places like the Chrome browser and OpenSSH.}

\stepcounter{footnote}\footnotetext{ Qubits and gates are randomly assigned.}

\subsection{Experimental Setup}
All compilation experiments are carried out on Intel Core i7-3960X (3.3GHz, 64GB RAM), implemented in the quantum compiler framework ScaffCC \cite{scaffcc} version 4.0. Noise simulations use Intel E5-2680v4 (28-core, 2.4GHz, 64GB RAM), performed using the IBM \texttt{Qiskit} software \cite{Qiskit}. Table \ref{tab:config} lists the ancilla reuse algorithms in our study. \emph{Eager} and \emph{Lazy} are two baselines that appear commonly in prior work. \emph{SQUARE} is our Strategic QUantum Ancilla REuse algorithm.

 \begin{figure}[t!]
    \centering
    \includegraphics[width=0.22\textwidth]{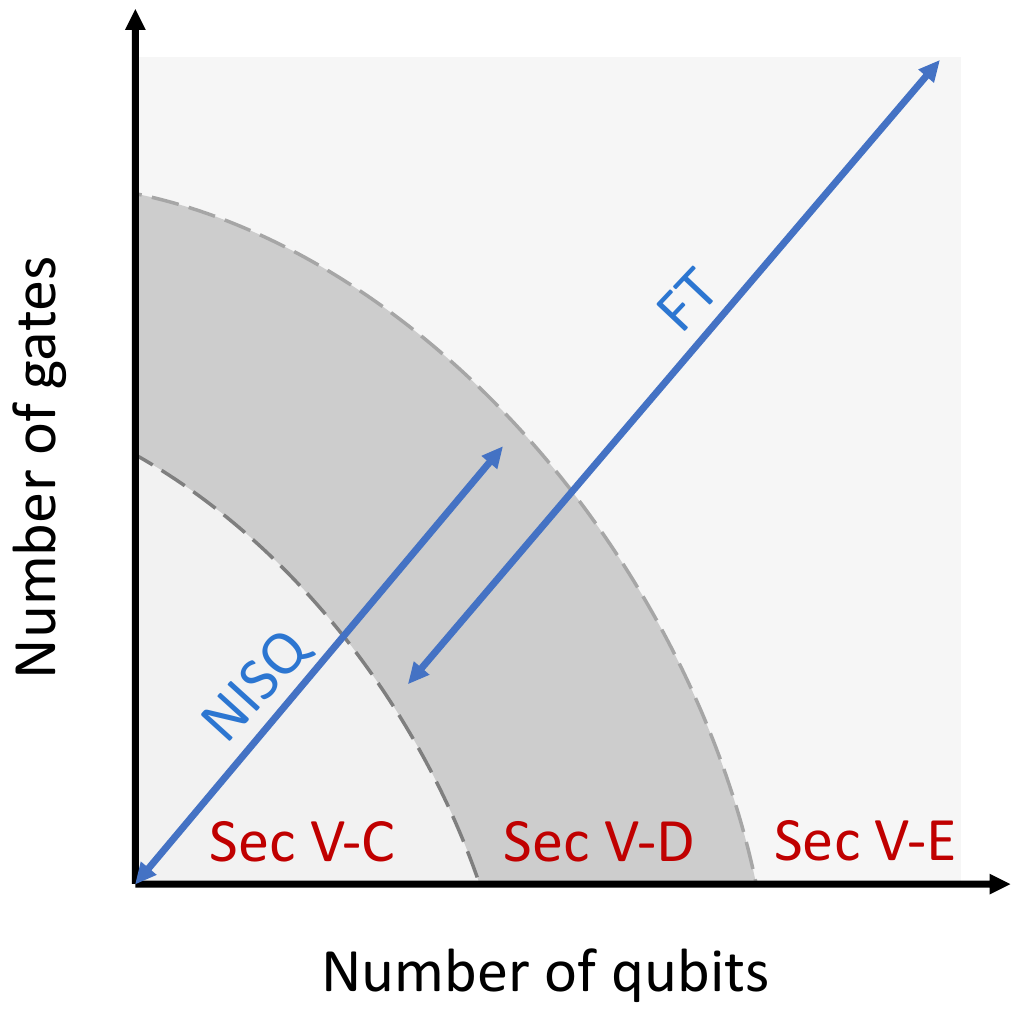}
    \vspace{-0mm}
    \caption{QC architecture boundary.}
    \label{fig:nisq_ft_boundary}
\end{figure}


The rest of the section are divided up into three main parts (Figure~\ref{fig:nisq_ft_boundary}) -- experimental results on NISQ architecture (Section \ref{subsec:nisq_exp}) with 2D lattice of physical qubits and nearest-neighbor connectivity as commonly used in \cite{IBM, holmes2018impact, shafaei2014qubit},  on NISQ-FT boundary architecture (Section \ref{subsec:nisqft-exp}) with same architecture model but on larger benchmarks, and on FT architecture (Section \ref{subsec:ft_exp}) with surface code error corrected logical qubits \cite{barends2014superconducting,ding2018magic}.

\begin{figure*}[!t]
    \centering
    \subfloat[Active quantum volume. (Lower AQV is better.) \label{fig:nisq_aqv}]{\includegraphics[width=0.32\textwidth]{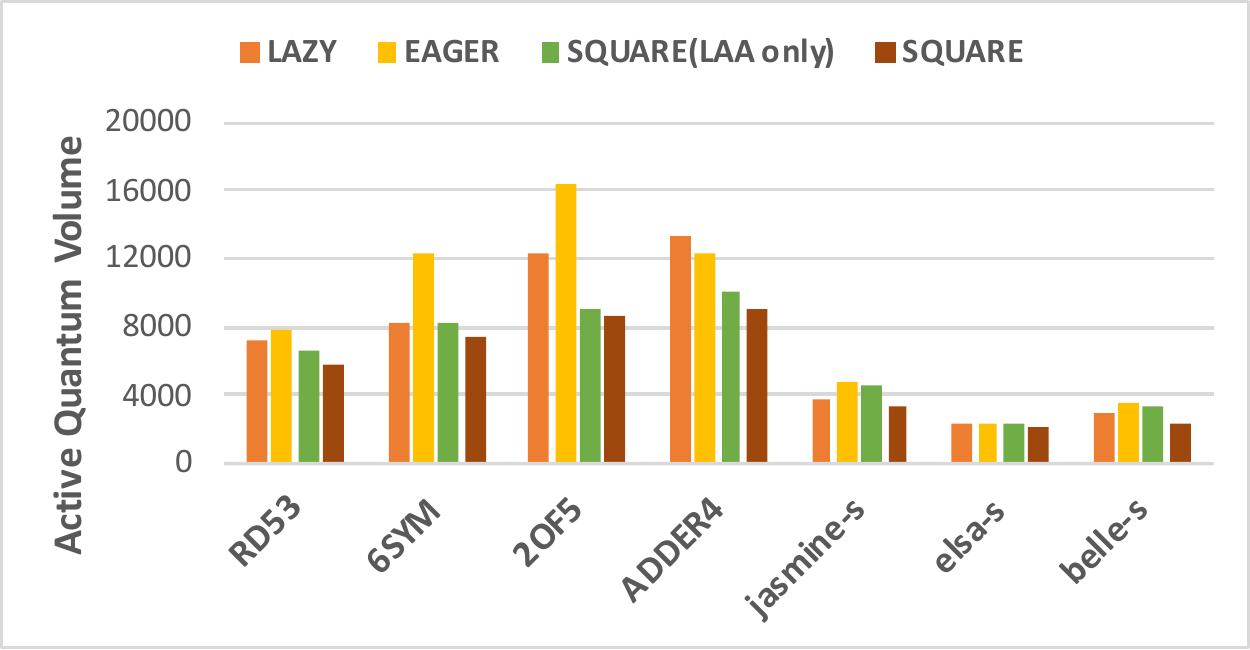}}
    \;
    \subfloat[Worst-case analytical model. (Higher success rate is better.) \label{fig:nisq_sr}]{\includegraphics[width=0.32\textwidth]{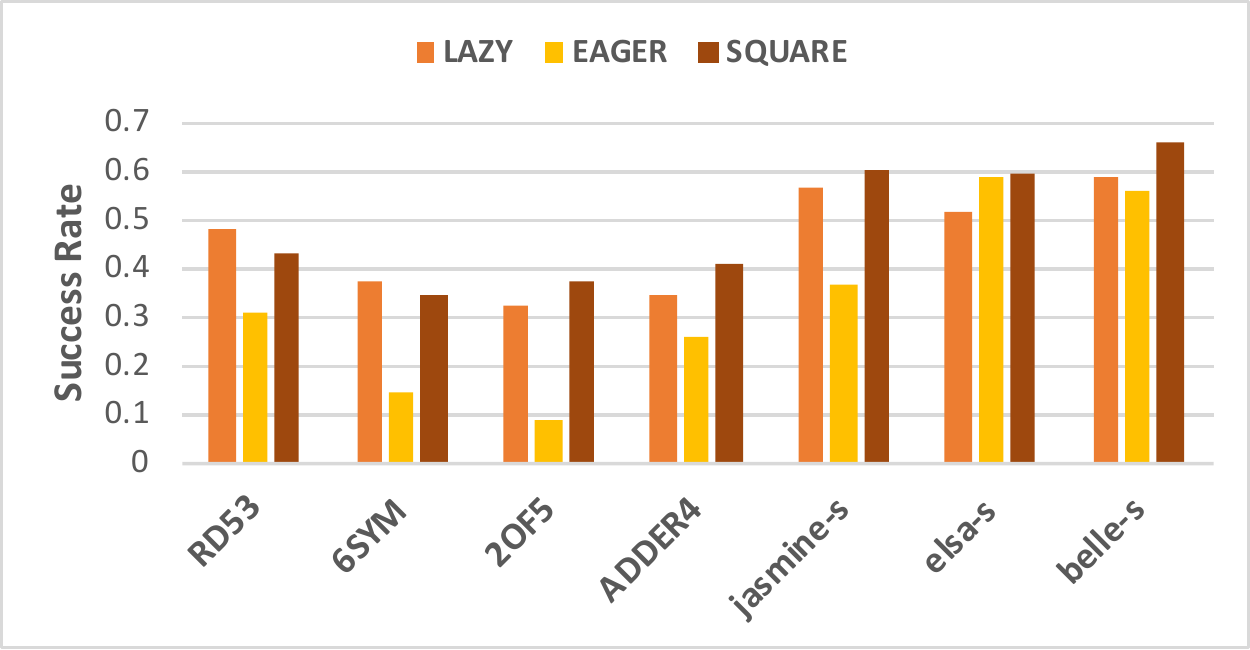}}
    \;
    \subfloat[Realistic noise simulation using IBM \texttt{Qiskit Aer} simulator. (Lower total variation distance is better.) \label{fig:nisq_sim}]{\includegraphics[width=0.32\textwidth]{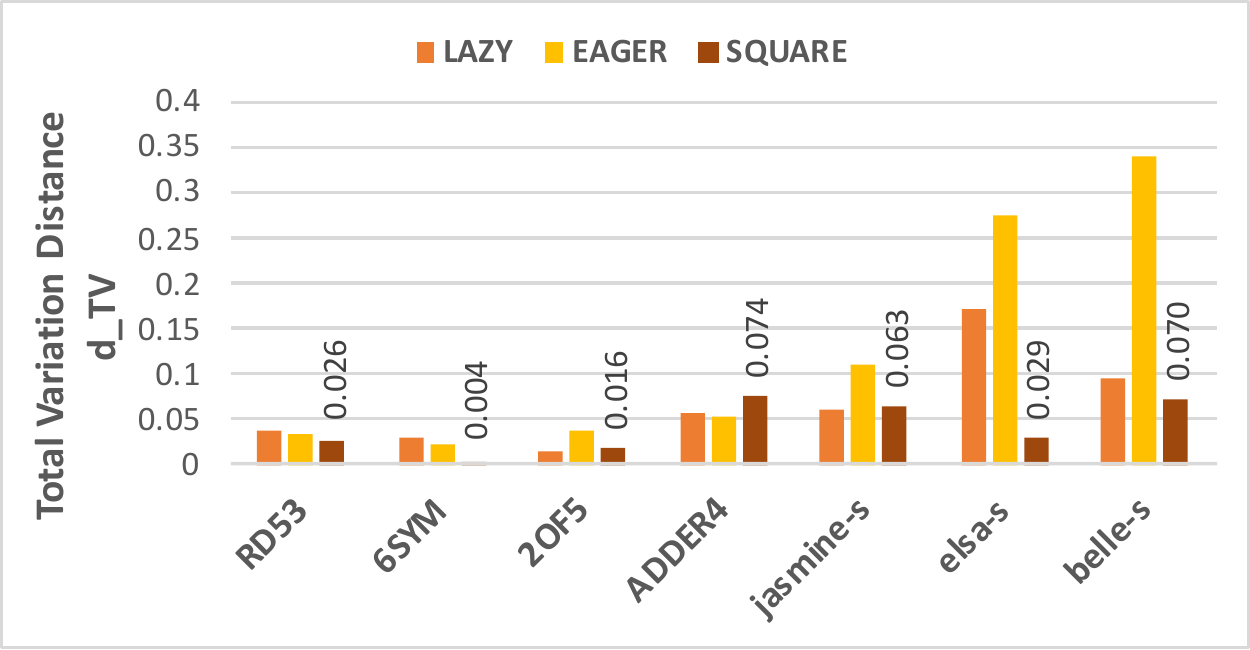}}
\caption{Impact of SQUARE optimizations on NISQ applications. All benchmarks use fewer than 20 qubits; SQUARE stands out as a strategy that uses substantially fewer qubits while maintaining high application success rate.}
\label{fig:nisq_results}
\end{figure*}

\subsection{NISQ Experiments}\label{subsec:nisq_exp}
Although SQUARE was initially designed to improve the performance of large-scale applications,
we find that reclaiming ancilla reduces program footprint and thus swap count due to communication on NISQ machines.  In this section, we give analytic and noise simulation results that quantify the fidelity gains from this reduced swap count.  To make noise simulation tractable, we focus on small benchmarks and introduce small versions of our 3 synthetic benchmarks as in Table \ref{tab:benchmarks}.


\subsubsection{AQV Analysis}

For our NISQ benchmarks, Figure \ref{fig:nisq_aqv} and Table~\ref{tab:nisq_results} show the characteristics of the compiled QC programs with different compiling policy. With the Eager compiling policy, the programs use the fewest qubits, but it may cost too many gates to reuse the ancilla qubits. SQUARE finds the balance between qubit uses and gate costs. We show the AQV comparison in Figure~\ref{fig:nisq_aqv}. The AQV is reduced when we apply LAA that allocates the closest qubits, reducing the number swaps. When full SQUARE is applied, AQV is further reduced because of reduction in uncompute cost.

\begin{table}[t!]
\scriptsize
\centering
\begin{tabular}{cccccc}
\hline\hline
Benchmarks  & Policy &\# Gates$^a$  &  \# Qubits &  Circuit Depth &  \# Swaps \\
\hline
&	Lazy &	536	 &	19	 &	395	& 	462	 \\
RD53&	Eager &	1064	 &	10	 &	878	& 	633	 \\
&	SQUARE &	932	 &  11	 &	635	& 	370	 \\
\hline
&	Lazy &	648	 &	19	 &	456	& 	654	 \\
6SYM&	Eager &	1293	 &	11	 &	1279	& 	1247	 \\
&	SQUARE &	1078	 &  12	 &	731	& 	520	 \\
\hline
&	Lazy &	708	 &	18	 &	723	& 	759	 \\
2OF5&	Eager &	1410	 &	8	 &	2374	& 	1728	 \\
&	SQUARE &	1176	 &  10	 &	952	& 	385	 \\
\hline
&	Lazy &	656	 &	18	 &	787	& 	725	 \\
ADDER4&	Eager &	1184	 &	12	 &	1139	& 	748	 \\
&	SQUARE &	920	 &  14	 &	715	& 	421	 \\
\hline
&	Lazy &	275	 &	16	 &	232	& 	73	 \\
Jasmine-s&	Eager &	1226	 &	5	 &	1055	& 	327	 \\
&	SQUARE &	510	 &  8	 &	427	& 	128	 \\
\hline
&	Lazy &	163	 &	15	 &	787	& 	725	 \\
Elsa-s&	Eager &	501	 &	8	 &	438	& 	163	 \\
&	SQUARE &	254	 &  13	 &	223	& 	85	 \\
\hline
&	Lazy &	220	 &	14	 &	202	& 	69	 \\
Belle-s &	Eager &	712	 &	6	 &	574	& 	113	 \\
&	SQUARE &	294	 &  9	 &	266	& 	89	 \\
\hline\hline\\
\end{tabular}
\footnotesize{$^a$ Here \# Gates does not include swap gates (listed in a separate column).}
\caption{NISQ benchmarks compilation results.}
\label{tab:nisq_results}
\end{table}

 \begin{figure*}[t!]
    \centering
    \includegraphics[width=0.68\textwidth, trim=0 0 0 0]{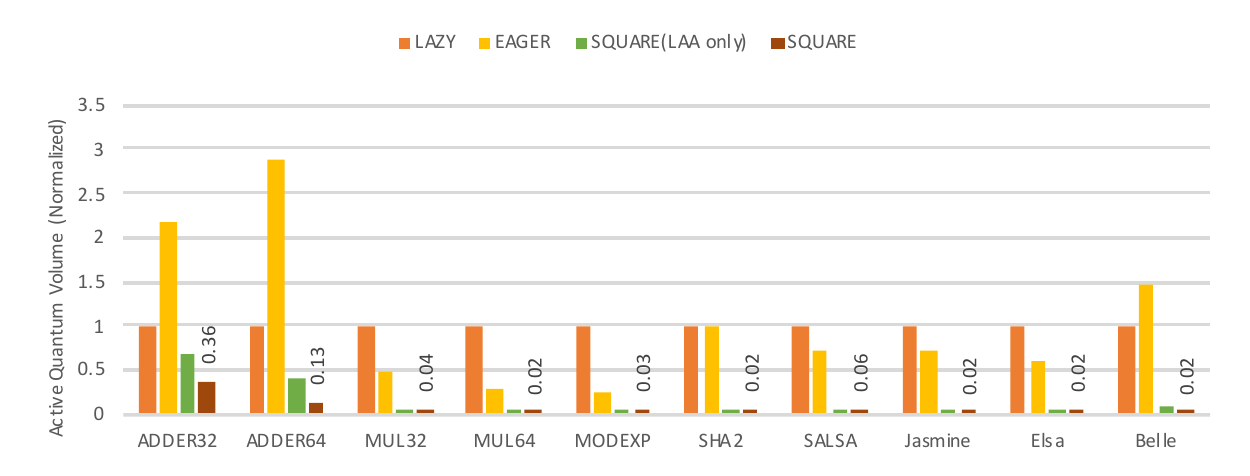}
    \vspace{-3mm}
    \caption{AQV results on medium-scale non-error-corrected quantum systems. Numbers on the chart correspond to the normalized AQV values of the SQUARE algorithm.}
    \label{fig:nisq_ft_aqv}
\end{figure*}

\begin{figure*}[t!]
    \centering
    \includegraphics[width=0.68\textwidth, trim=0 0 0 0]{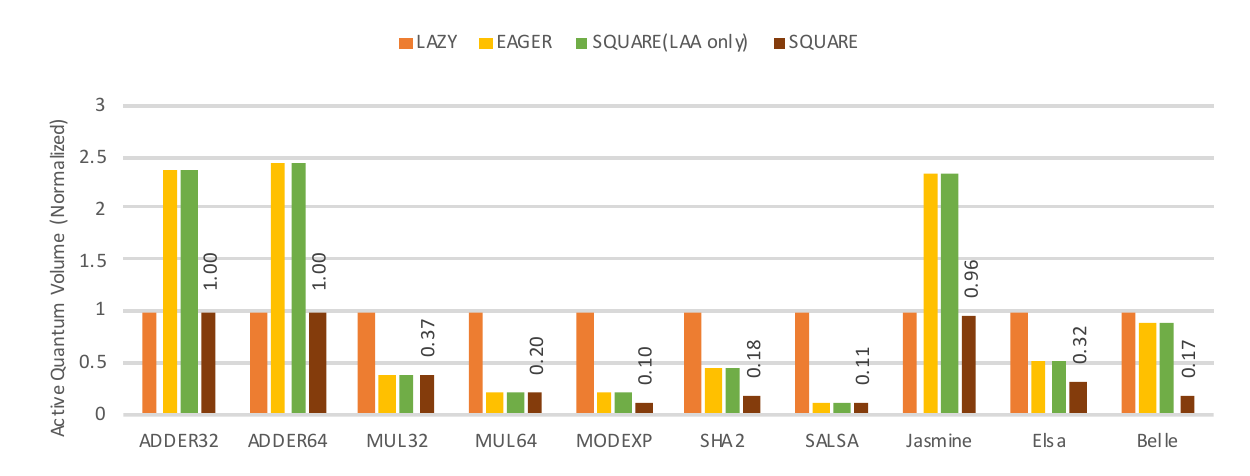}
    \vspace{-3mm}
    \caption{AQV results on fault-tolerant quantum systems.} 
    \label{fig:ft_aqv}
\end{figure*}

\subsubsection{Program Success Rate by Analytical Model}
Program success rates in our evaluation are estimated by a worst-case analysis using qubit decoherences and gate errors. Multiplying the single-qubit/two-qubit gate success rates and the probability of qubit coherence from Table \ref{tab:error_rate}, we observe an average improvement by 1.47X w.r.t Eager and 1.07X w.r.t. Lazy. With strategic uncompuptation by SQUARE, programs use fewer qubits and improve overall chance of success. In reality, this worst case analysis may neglect program structures and noise cancellation.  Results are even more positive in the next section where we perform noise simulation.

\begin{table}[t!]
\footnotesize
\centering
\begin{tabular}{ccccccc}
\hline\hline
  & \# Qubits  & $\epsilon_{single}$ & $\epsilon_{two}$ & T1 ($\mu s$) & T2 ($\mu s$)\\
\hline
IBM-Sup \cite{IBM, IBMcalibrate} & 	20	 &	$<1\%$	 &	$<2\%$	& $55$ & $60$\\
\hline
IonQ-Trap \cite{ionq} &	79	 &  $<1\%$	 &	$<2\%$	& $>10^6$ & $>10^6$\\
\hline
Our Simulation &		$<20$	 &	$0.1\%$	 &	$1\%$	& $50$ &	$70$ \\
\hline\hline
\end{tabular}
\caption{Error rates on real devices and noise models on our simulation.}
\label{tab:error_rate}
\end{table}

\subsubsection{Noise Simulations}
All simulations in our evaluation use IBM \texttt{Qiskit Aer} simulator \cite{Qiskit} with noise models from the \texttt{qiskit.providers.aer.noise} library -- \texttt{depolarize\_noise} for single-qubit and two-qubit gate noises, and \texttt{thermal\_relaxation} for T1/T2 relaxations to account for qubit decoherence. Table~\ref{tab:error_rate} shows the parameters in our simulation, compared against those in real devices. Figure \ref{fig:nisq_sim} shows the results from simulation; each data point is obtained from 8192 shots of noisy circuit simulation. We use total variation distance $d_{TV}$, to compare measurement outcomes of noisy circuits with those of ideal ones; it's a common measure for QC experiments \cite{buadescu2019quantum, lund2017quantum, arute2019quantum}. We observe that SQUARE achieves lowest distance for almost all benchmarks compared to Eager or Lazy. 

\subsubsection{Applicability of SQUARE to NISQ Machines}
Table \ref{tab:nisq_results} and Figure \ref{fig:nisq_sr} together show the impact of uncomputation on circuit fidelity. SQUARE finds a balanced middle-ground between qubit savings and gate costs by strategically uncomputing its functions. Surprisingly, when comparing Lazy with SQUARE, the additional gates for uncomputation \emph{reduces} the total number of operations, thanks to a substantial reduction in swap gates, as ancilla qubits with better locality are actively reclaimed and reused. Uncomputation also dis-entangles garbage qubits from output qubits, preventing noise from propagating. Furthermore, SQUARE retains most of the qubit savings as Eager does. Overall, SQUARE achieves high success rate using fewer qubits than Lazy.

\subsection{NISQ-FT Boundary Experiments}\label{subsec:nisqft-exp}
The boundary between NISQ and fault-tolerant architectures are far from clear. For completeness, we analyzed the performance of the SQUARE algorithm assuming medium-scale machines (with 100-10000 qubits) is built without error correction. 
Figure~\ref{fig:nisq_ft_aqv} shows the AQV results with different compiling policies, and the normalized AQVs of SQUARE are labeled. We observe significant AQV savings by SQUARE, reducing the AQV by a factor of 6.9X on average when compared to the Lazy policy.

\subsection{Fault-Tolerant (FT) Experiments}\label{subsec:ft_exp}

The FT experiments share the set of benchmarks used in the NISQ-FT experiments, but use braiding for communication. To do so, we build and integrate a braid simulator in SQUARE to precisely calculate the communication overhead for executing a program on a surface-code error-corrected architecture.

Following prior work \cite{fowler2012surface, javadi2017optimized, ding2018magic}, we assume logical qubits on the surface are laid out in a 2-D array, with sufficient distance between qubits. The separation between qubits serves as channels, allowing other qubits to braid through. So in our simulator, we associate one site per qubit and channels wide enough for a single qubit to braid through. Furthermore, different single-qubit gates have different time cost. 

We substitute the swap-chain generation procedure in the SQUARE's gate scheduler with a braid generation procedure. In particular, when a {\tt CNOT} gate is scheduled, we first find a route between the operand qubits, and then check if it crosses with other ongoing braids. It is queued until its route has been cleared.

As shown in Figure~\ref{fig:ft_aqv}, SQUARE significantly reduces AQV in all applications under the FT system environment. Comparing to Lazy policy, SQUARE achieve 44.08\% AQV reduction on average, and up to 89.66\% reduction.

%% file: related.tex

%% file: conclusion.tex
\section{Conclusion}\label{sec:conclude}

We have presented an automated compilation tool flow
 that manages the allocation and reclamation of qubits in reversible program with modular structures. We choose a dynamic heuristic-based approach to tackle the challenges, proving how we can use the knowledge of qubit locality and program structure to our advantage to efficiently compile high-level arithmetic for a resource-constrained machine. That is accomplished by 
 SQUARE via cost-effective uncomputation. In this process, we introduce a resource metric, AQV, that quantifies the amount of resource utilized by a given computational task. It allows us to measure and compare the effectiveness of various compiler optimization designs.

The core of our optimization tool flow is the allocation and reclamation heuristics, which predict the cost of uncomputation based on information such as qubit savings, gate overheads, potential reuse, and decisions in children modules in the program call graph. Our methodology is shown to be effective on a suite of benchmarks, including common arithmetic functions and synthetic programs with arbitrary structures. We evaluate SQUARE on NISQ systems and FT systems. The results show that our study has practical value for not only current NISQ devices but also future FT systems.

%% file: main.bbl
\begin{thebibliography}{10}
\bibitem{IBMcalibrate}
``Coherence times are obtained from daily calibration data from ibm q
  experience.''
\end{thebibliography}
